\documentclass[aps,twocolumn,groupedaddress]{revtex4}
\usepackage[latin1]{inputenc}
\usepackage{amsmath}
\usepackage{verbatim}
\usepackage{amssymb}
\usepackage{bbm}
\usepackage{graphicx}
\usepackage{multibib}

\begin{document}

\title[]{Mammalian cochlea as a physics guided evolution-optimized hearing sensor}
\author{Tom Lorimer, Florian Gomez, Ruedi Stoop}
\email{ruedi@ini.phys.ethz.ch}
\affiliation{
Institute of Neuroinformatics, University of Zurich and ETH Zurich, Winterthurerstrasse 190, 8057 Zurich, Switzerland}

\date{\today}

\begin{abstract}
Nonlinear physics plays an essential role in hearing, from sound signal generation to sound sensing to the processing of complex sound environments. We demonstrate that the evolution of the biological hearing sensors demonstrates a dramatic reduction in the solution space available for hearing sensors due to nonlinear physics principles. More specifically, our analysis hints at that the differences between amniotic lineages hearing, could be recast into a scaleable and a non-scaleable arrangement of nonlinear sound detectors. The scalable solution employed in mammals, as the most advanced design, provides a natural context that demands the ultimate characterization of complex sounds through pitch.
\end{abstract}

\maketitle


\section{Introduction}
Nature provided our planet with an abundance of species. The question of how this abundance comes about has intrigued humans since early in their existence. In his treatise `On the Origin of Species', Charles Darwin set forth in 1859 \cite{Darwin} for a scientific explanation, anchoring it in the general principles of competition. Since then, research on evolution has focused mostly on the particular twists and turns the course of natural selection has taken, trying to understand what advantage a specific modification would have given to its bearer. While looking at the direction evolution has taken, we observe in a number of instances an apparent convergence towards certain building principles that come unexpectedly. The mammalian ear is one of these examples.

\begin{figure} [h!!!!!!]
\centering
\includegraphics[width=1\linewidth]{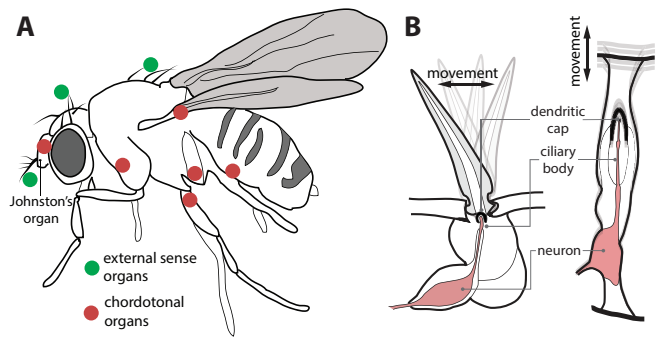}
\caption{Sensory hair cells and chordotonal organs \cite{Drosophila_genes}.  A) Locations of sensory hair cells (including the antennal receiver) and chordotonal organs in {\it Drosophila}. B) Insect variant of inner and outer hair cells (arista and chordotonal organs), based on common genetic origins \cite{WeberPNAS}.}
\label{lizards1}
\end{figure}

After a long tradition of research on evolutionary linkage \cite{c1,c2,c3,c4,c5} and on physiological and genetical correspondences of species \cite{manleylizards, d1, d2, d3, d4, d5, yack}, it was proposed that a convergent evolution may have directed insect \cite{c6}, as well as jointly insect and mammalian audition \cite{Montegalegre}. 
Hearing in both cases is mediated by the same key genes \cite{Boekhoff}, which indicates a close evolutionary relationship. In mammalian audition, the anion transporter family {\it prestin} is expressed, whereas audition is mediated in nonmammalian vertebrates and in insects by prestin-homologous proteins \cite{WeberPNAS}. 
The chordotonal organ (e.g. in Johnston's Organ of the mosquito or of \emph{Drosophila}  \cite{yack}, c.f.  Fig. \ref{lizards1}), provides sensory basis of most insect hearing. Although seemingly very different at first view, the human cochlear hair cells that we will later centrally deal with, follow genetically closely the building principle of the chordotonal organs \cite{Drosophila_genes}. All of these points have been interpreted to hint at a joint early origin and parallel evolution of the hearing system.

Whereas these approaches have shed a fascinating light on how a major biological sense evolved and developed, they cannot provide a full understanding of what exactly has kept similar genetic origins on this parallel, or as we shall claim: even convergent sensory evolution. Moreover, blueprints for constructing artificial sensing devices of a quality matching those of the biological sensors (which often is a motivation behind such an endeavor, and which is often considered as the proof of a full understanding of the matter), should not be expected from genetical or physiological analysis. This seems to require to understand the role that physics could have had in the evolution of hearing. 

Here, we investigate what constraints macroscopic physics (in contrast to the genetical or biochemical microscopic views) imposes on the construction of an optimized hearing organ. Exclusively on this level, we will exhibit how physical principles constrain the solution space of optimal hearing sensors in such a way that for large frequency bands and sharp resolution hearing, a convergence towards the blueprint realized in the mammalian cochlea is highly likely to occur. Our implicit claim is that the observed parallel and convergent evolution of the biological hearing sensors is the result of physical constraints that are at work during the evolution process, rather than a joint genetic origin. We will, finally, argue that the optimized solution realized in the mammalian cochlea naturally entrains the notion of 'pitch sensation' as experienced by humans. 

\section{Small-power single frequency sensing} 
We start by positing that sounds around a dominant frequency are of particular interest to the animal world (the question how periodic behavior emerges from complex entities such as animals is old; if desired, the reader will find an appendix indicating our view regarding this issue). For spotting a predator, or a conspecific for reproduction, hearing a weak sound first among competitors is a substantial evolutionary advantage. In the simplest case, identifying one characteristic frequency will be important and might be sufficient. Insect hearing illustrates this: The male mosquito \emph{Aedes aegypti} performs `near-field'' hearing with a sensor that is tuned to the wingbeat frequency of females \cite{goepfert_mosquito}. For this, and to exhibit the close relationship to human hearing for single frequency hearing, we will use in this section the well-understood insect example for the illustration of our arguments.

\begin{figure}[h!!!!!!!!!!!!!!!!!!!]
\centering
\includegraphics[width=1\linewidth]{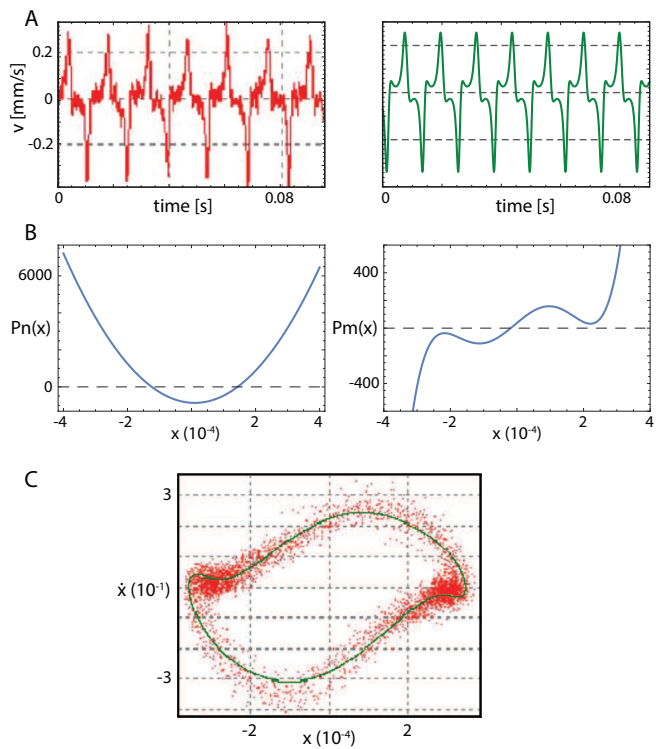}
\caption{Fully developed self-sustained antenna oscillations (SO) of Drosophila, 20 min after DMSO 
injection (after Ref. \cite{Stoop_Smirnov}: A) Red: Data, green: simulations.
B) Best data-based polynomial approximating ordinary differential equation of SO $\ddot{x}+Pn(x)\dot{x}+Pm(x)=0$, with $n=2$ and $m=5$, at extracted parameters this system is close to a Hopf bifurcation (the classical Hopf equation $\ddot{x}-\mu(1- x^2)\dot{x}$ has a Hopf bifurcation at Hopf parameter $\mu=0$). C) Original (red) and low-passed reconstructed (green) SO data. 
\label{Smirnov}
}\end{figure}

For sound detection and perception, very faint input level sounds first need to be amplified actively (i.e., by using energy added from outside), which then will enable the processing of this information on a fully developed level. For the amplification, a quite general physical principle is used that is very deeply rooted in physics; how it is effectively implemented (e.g. whether on a molecular, mechanosensitive or electromotile level) is for the present discussion actually irrelevant. Bifurcation theory developed in mathematics thirty years ago dealt with the fact that if in physical systems parameters are changed, occasionally the solutions emerging from such systems change their nature \cite{Hopf1}. By varying a parameter across a certain value (the so-called bifurcation point), the nature of solution changes, in many cases by going from rest into an oscillatory state. Close to the bifurcation point, the natural solution loses its stability, and small perturbations develop in a hardly controlled manner, until after a generally considerable time lapse, the system settles on its natural solution. The closer a system approaches instability introduced by the bifurcation, the more small inputs to the system are converted by the system into huge responses. In this way, systems close to bifurcations have been proposed to be used as active small-signal amplifiers \cite{Wiesenfeld,Stoopbrun}. 

Two prominent bifurcations \cite{Hopf1} are generic candidates for the required bifurcation: a saddle-node (tangent) bifurcation (such as that leading from quiescence to regular spiking in the neuronal Morris L\'{e}car equations) or a Hopf bifurcation \cite{Hopf} (as found in the  Hodgkin-Huxley axon equations). While both bifurcations may serve as small signal amplifiers, the particular bifurcation delivers a specific fingerprint onto the amplification law, which in insect  case considered below, as well in human hearing \cite{KernStoop_prl_2003,KernStoop_pnas,KernStoop_prl_2004} points at a Hopf bifurcation as the relevant process. 

\begin{figure}[h!!!!!!!!!!!!!!!!!!!!!!!!!!!]
\centering
\includegraphics[width=1\linewidth]{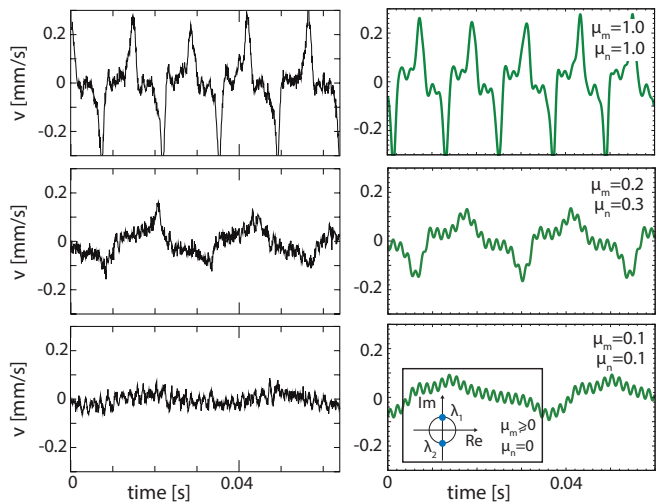}
\caption{Top to bottom: From self-sustained antenna oscillations back to the quiescent fixed-point. Left column: experimental data, right column: simulation, where the polynomials were reduced by factors $\mu_m\simeq \mu_n$.
Close to bifurcation, $\mu_n$ precedes  $\mu_m$, so that at bifurcation $\mu_m>0$. Inset: At crossing to quiescence, the linear analysis reveals a Hopf bifurcation.}
\label{fig:aplnova_fig55}
\end{figure}

Evidence for a Hopf bifurcation underlying the amplification process is in the insect case obtained as follows. Biological small-signal amplifying systems rest typically below the bifurcation point to oscillation. The bifurcation point may, however,  even be crossed under certain conditions, which can be used to infer the deeper nature of the active amplification process below bifurcation. In the example of the Drosophila antenna \cite{Stoop_Smirnov}, an injection of  biochemical dimethyl sulfoxide (DMSO) leads to a crossing of the bifurcation point, from stirred antennal vibrations to self-sustained oscillations \cite{Stoop_Smirnov}. 
From the observed velocity time series of the antenna oscillations (Fig. \ref{Smirnov}A), an underlying generalized van der Pol system (equation: see caption Fig. \ref{Smirnov}) could be identified that operates in the close vicinity of a Hopf bifurcation. For expressing the short-scale oscillations, a term $A_0 \cos(2 \pi \, f \, t)$ was included into the equation ($A_0 = 70$ and $f=600$ Hz). This term does not compromise the nature of the bifurcation and can be omitted for the following discussion.
An enlightening understanding of the amplification dynamics can be provided by the behavior around zero displacement position $x = 0$, where the nonlinear damping term $Pn(x)<0$ implies that energy is injected into the system, indicating active amplification (Fig. \ref{Smirnov}B). Around $x=0$, the nonlinear restoring force $Pm(x)$, together with its first and second derivatives, are relatively small. This implies that for small receiver displacements, virtually no restoring force is present. By means of the negative damping term, the system is thus easily driven out to large amplitudes. The comparison between data and obtained trajectories reveals the close correspondence between the data and the model. 
After having evaluated the system equations for the fully self-sustained oscillatory state, we follow the system on the way back to below the bifurcation point (Fig. \ref{fig:aplnova_fig55}).
The recorded data compared to the figures obtained from scaling the two polynomials by two factors $\mu_m, \,\mu_n$, demonstrate, that we follow closely the biological changes on this way, where $\mu_n$ first lags somewhat behind $\mu_m$, but then takes the lead, so that $\mu_m$ is still positive at bifurcation point, which is where the linear analysis reveals a Hopf bifurcation (inset).
Below, but close to the bifurcation point, where the antennal system usually operates, system-specific details are drowned out by the bifurcation properties. This implies that any such system, in particular Drosophila's antenna equations, can be described in its essential features by the prototypical Hopf equation \cite{Hopf}.

\begin{figure}[h!!!!!!!!!!!!!!!!!!!!!!!!!!!]
\centering
\includegraphics[width=1\linewidth]{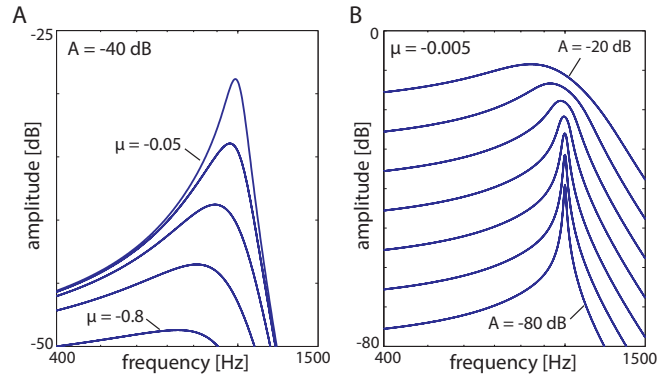}
\caption{Single Hopf amplifier response \cite{KernStoop_prl_2003} conditioned on the passive behavior in the cochlea (leading to the asymmetry if compared to \cite{Wiesenfeld,Stoopbrun}). The description mimicks the behavior of outer hair cells with a preferred frequency CF embedded into the basilar membrane: Frequency selectivity A) regarding different distances $\mu \in \{-0.05, -0.1, -0.2, -0.4, -0.8 \}$ from bifurcation point, B) regarding input signal strength (increase in steps of 10 dB).}
\label{fig:aplnova_fig5}
\end{figure}

Discounting microscopic differences, not just the solution construction, but also the generic function principles are shared between insects and mammals. We exhibit how these actually work, by looking at the example of the Hopf system that has been shown to be at the heart of  human hearing. In this case, the nonlinear amplification is by the electromotile outer hair cells embedded into the basilar membrane, working in the vicinity of, but below a Hopf bifurcation. If stimulated by a signal of frequency $\omega$ close to the Hopf system's characteristic frequency $\omega_0$, the system would oscillate at $\omega$, at a considerable amplitude. The response shown in Fig. \ref{fig:aplnova_fig5} embraces all the required amplification properties of a small signal amplifier. It is worth noting that these amplification profiles are of fundamental importance; we will show that their properties are preserved the whole way up the auditory pathway.  From them, the main properties of the mammalian hearing sensor can be reproduced and understood (\cite{MartignoliStoop2010,MartignoliGomezStoop2013}, in particular the supplemental materials). The outer hair cells in today's cochleae emerged very early in evolutionary history, before even the split of the stem reptiles from which the amniotes evolved, approximately 400 million years ago \cite{manley}. Why, how and under what conditions they came to form the final mammalian hearing sensor, the cochlea, is the focus of the next section.

\section{Many-frequency challenge} 
Beyond the primary task of identifying a weak characteristic frequency, the distinction of several frequencies emitted by a mate, predator, and prey, may be of importance, for procreation and survival,
in species that interact more intricately with the world around. 
 For a scheme to evolve to amplify and distinguish a broad suite of frequencies, these hair cells must somehow embody a tuning mechanism. 
The simplest solution on first view would be a construction scheme by which each sensor would inherently react to one particular frequency. 
One complication, however, emerges in this case: The superposition principle does not hold for nonlinear amplifiers. Together with the desired frequencies, also nonlinearity-generated undesired families of interaction products among the amplifiers emerge. To suppress the amplification by adjacent amplifiers with characteristic frequencies matching the combination tone frequencies, the sensors would need to be `well-separated', e.g. by placing them at sufficient distance from one another. Since this also entails long wiring, such an arrangement will be preferable where relatively few frequencies are to be dealt with, as is often the case for small-sized animals, such as insects. In fact, the chordotonal organs upon which insect hearing is based, are found all over the insect body (Fig. \ref{lizards1}). From these separated sensors, a huge variety of more specialized ``hearing'' organs has developed in different parts of the anatomy for different species \cite{yack}. 

\begin{figure} [hhhhh!!!!!!]
\centering
\includegraphics[width=1\linewidth]{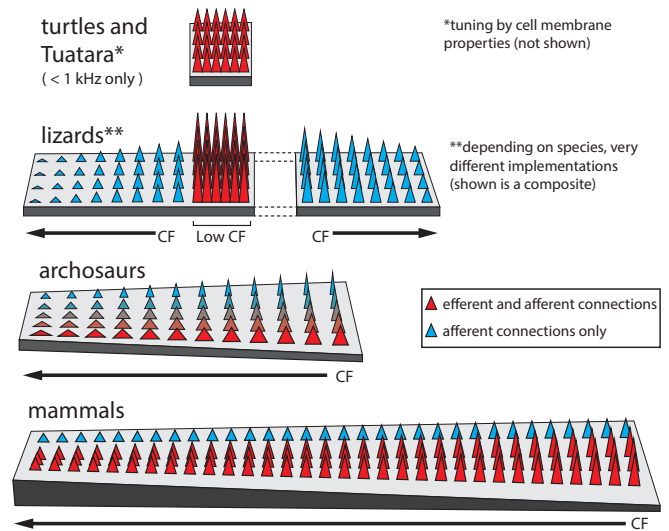}
\caption{Schemes of frequency tunings and spatial arrangements across stem reptile descendants. Shown are hair cell/bundle morphological changes (height/width), basilar membrane changes (as relevant to tuning), and hair cell innervation changes (blue: virtually none, red: increased efferent innervation).  Hair cells membrane changes (electrical tuning), orientation, relative numbers of hair cells are not reflected. Short, unspecialised basilar papilla of turtles and Tuatara populated by single type of electrically tuned hair cells sensitive only to low frequency.  Lizard families separate high- and low-frequency areas on (modular) ``untuned'' basilar papilla using different hair cell types. A tectorial membrane may be present/absent.  Archosaurs (mammals) implement gradually (fully), a single tonotopic gradient through basilar membrane stiffness and surface tension, combined with a hair cell length scaling that can be controlled at all frequencies by efferent innervation (blue cells: inner hair cells, red cells: outer hair cells).  
}
\label{lizards2}
\end{figure}

For higher animals such as mammals, a similarly distributed hearing systems would, however, increase the expense and complexity of connection and integration. The solution is evidently to locally concentrate the sensors and to live with the emergent complexity.
The simplest organization of the hearing sensors in the stem reptile descendants is found with the turtles and Tuatara. Their characteristic frequencies are electrically implemented and they have generally a very limited range of frequency sensitivity implemented in a relatively unsophisticated sensor arrangement (Fig. \ref{lizards2}). From an evolutionary optimization viewpoint, their solution seems to be still similar to the frequency tuned hearing organs of insects, by which much of the complexity in the auditory environment is not probed. 

The lizards, as the next step of hearing sophistication, show morphological gradients and variations of hair cells (Fig. \ref{lizards2}), influencing the frequency tuning which is no longer purely electrical, but also (micro-)mechanical. Their sensor has enlarged its range of accessible frequencies, which entails an increased complexity of the auditory signals they have to cope with. 
Under such conditions, among the hair cells, the emergence of strong irregular combination tones is natural, and to deal with them will have placed a significant cognitive burden higher up in the auditory pathway. As a result, the great architectural variety of the lizards' sensors could be interpreted as local minima during the optimization process to cope with this difficulty, but where evolutionary pressure has not yet pushed the constructions into the global optimum. Compared to mammals and birds, lizards base their living on auditory information to a lesser extent; they might have minimized the effect of the interaction products between sensors of different characteristic frequencies at a price of much reduced hearing discrimination. Similarly, lizards have developed two distinct kinds of hair cells and  separate them into type-specific areas, but only the low-frequency kind is provided with efferent connections that enable a neural frequency tuning of these elements \cite{manleylizards}. This reduces the 'listening' capability of the sensor as a whole \cite{Stoop_GomezREAPPL} (the tokay gecko, which uses two types of hair cells similar in character to the mammalian inner and outer hair cells, may be seen as the exception to the general rule \cite{Chiappe}).

For an extended frequency range such as is essential for mammals, it is of great importance to construct the sensor in the utmost efficient manner (how the complexity of the generated signal can be kept under control will be dealt with later). Tonally, this implies some broad structural arrangement of the characteristic (preferred) frequencies across the device. This might be implemented in one or two dimensions. We suggest that a single one-dimensional frequency arrangement (such as first observed in the archosaurs and mammals) can be considered as the starting point for efficient sensor design:  Frequencies require only one dimension for definition, so higher dimensional placements (such as concentric rows on a circular membrane) are likely to compromise spacing or wiring efficiency  (wiring efficiency is one of the driving organizational principles of Cortex \cite{stoop_prl_2013, Houzel_2010}). This seems to be the solution that independently archosaurs (the ancestors of birds and crocodiles) and, some millions of years later, mammals have converged to. Both lines developed an elongated sensor-substrating basilar papilla and two kinds of hair cells on it. In the archosaurs, the tuning of the sensors remained partially electrical, the mammals, however, got rid of this constraint that limits high-frequency hearing.

In the literature, a number of tuning ``parameters'' are cited to control the frequency sensitivity of hair cells. They distinguish broad scale and localized changes in the mechanical properties of the basilar papilla and/or tectorial membrane (macro- and micro-mechanical tuning), as well as local modifications to the properties of the hair cells themselves (micro-mechanical and electrical tuning) \cite{manley}. 
We believe that many of these observations are of primary physical rather than of genetical origin. 
As was exhibited, archosaurs and eventually mammals follow an essentially one-dimensional tonotopical placement of the hair cell sensors, where frequency and distance space collapse on a logarithmic scale. 
Outer hair cell frequency sensitivity is then essentially defined by membrane substrate stiffness and surface tension, and by hair-cell size. Indeed, investigations of the mammalian outer hair cells have revealed that a single hair cell is likely to be broadly tuned in isolation \cite{broadtuninginisolation}; its sharp frequency specificity is mostly obtained from its embedding into the basilar membrane as the carrier. From the basilar membrane side, frequency sensitivity is achieved by the exponential decrease of the basilar membrane stiffness and a corresponding modification of the surface tension \cite{disskern,KernStoop_prl_2003}.


\section{Scalability and consequences} 

\begin{figure}
\centering
\includegraphics[width=1\linewidth]{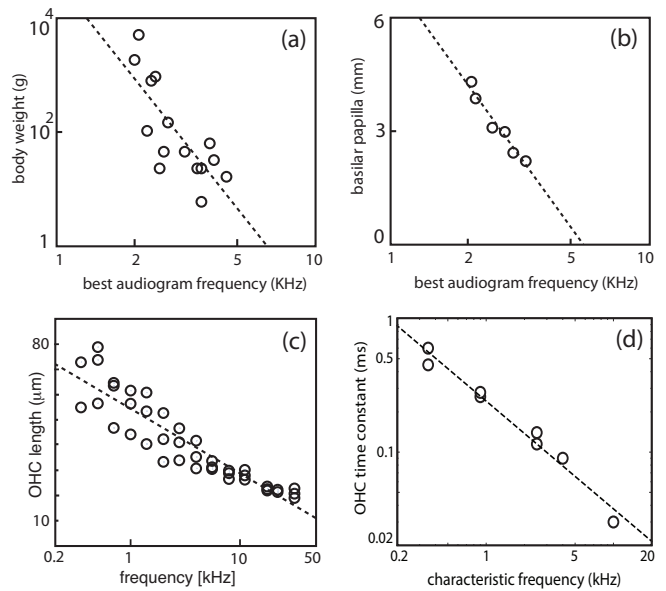}
\caption{Scaling: a) body weight, b) basilar papilla length vs. best hearing frequency  \cite{Gleich_2005}; c) outer hair cell length vs. characteristic frequency in situ (each circle refers to the mean from one guinea pig OHC row, based on data from Ref. \cite{pujol}), d) outer hair cell time constant vs. characteristic frequency.}
\label{fig:aplnova_fig2}
\end{figure}

The tonotopic organization finally laid the basis for scaling as the construction plan for the bird and mammal hearing sensors.
Scalability of the hearing sensor is important. It is important in the context of evolution of the species within a single family, where it is reflected in the emergence of approximate natural scaling laws between the properties of the originator of a sound and the sound itself. We observe, for instance that the relationship between the weight of an animal, and the frequency it hears best can be approximated by a power law (Fig. \ref{fig:aplnova_fig2}a).  Moreover, the frequency of best hearing is correlated with the high-frequency limit of hearing: small species with a short basilar papilla hear higher frequencies, compared to larger species with a longer basilar papilla \cite{Gleich_2005} (Fig. \ref{fig:aplnova_fig2}b). 

Whereas in the archosaurs frequency tuning in the hair cells is to a large extent due to electrical tuning \cite{manley}, this strategy was replaced in mammals by a tuning by means of the outer/inner hair cell size and basilar membrane change. With this, the mammalian hearing sensor finally achieves a scalable solution. As we go down the cochlear duct, the stiffness of the basilar membrane decays exponentially. To keep pace with this, outer/inner hair cell increase their length correspondingly (Fig. \ref{fig:aplnova_fig2}c), which entails compatible whole-cell slope conductances and capacitances \cite{Housley} (Fig. \ref{fig:aplnova_fig2}d).
The mammalian concept thus has the advantage that the frequency properties of each sensor don't need to be genetically set, but can be obtained from essentially scaling one single construction. Where the scalable construction offered by the elongation of the basilar papilla in non-mammalians was still limited by the electrical frequency tuning of the hair cells \cite{manley}, in the mammals, finally, electrical tuning was abandoned, which removed this limitation. 

The lacking of an low upper limit in frequency space led in mammals to an exquisite elongation of the basilar membrane, which then by the spiraling for space, then led to the mammalian cochlea's final form. Scalability is arguably also the most efficient way of overcoming the general wiring constraints with which the cortex is confronted. We have shown recently that the observed doubly fractal network architecture of the cortex, minimizes the network's total wiring length required to generate a coherent information wavefront at any given speed of information transfer  \cite{stoop_prl_2013}. It seems not too far-stretched to assume that the construction of the cochlea serves a similar constraint regarding its interfacing with the cortex (recent modeling experiments using quite arbitrary cochlear stimuli have, in fact, indicated a very stable scale-free avalanche size distribution of the `excited nodes' in the network \cite{inpreparation}).

\section{Complex excitation patterns} 
 A strong point of support for our thesis of physics guiding the evolution of the hearing sensor, is the observation that among all mammals, the sensor construction is quite uniform. The human cochlea, e.g., is extremely similar to that of a squirrel, cat, dog, or of a guinea pig.
  \begin{figure}[hhhhhhht]
\centering
\includegraphics[width=1\linewidth]{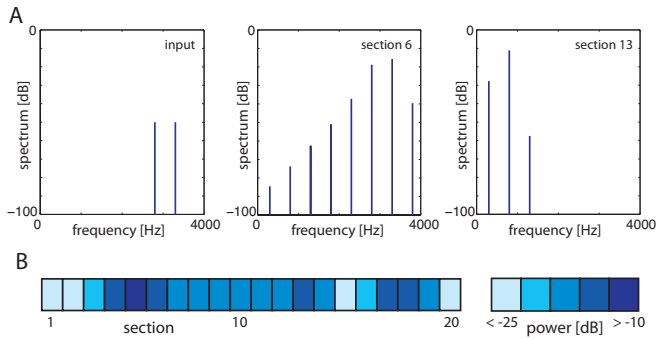}
\caption{Two-tone stimulation, -50 dB each tone, of the Hopf cochlea (7.04 - 0.22 kHz, 20 sections all tuned to $\mu=-0.1$).  
A) Signal change along a discretized cochlea  \cite{stoopcochlea1,stoopcochlea2,vyver}, showing the emergence of complex excitation patterns by the emergence of combination tones further down the cochlear duct. B) Generated cochlear excitation pattern.}
\label{fig:aplnova_fig6}
\end{figure}
In all these cases, the interactions of the small-signal amplifiers produce combination tones. As the sound travels down the basilar membrane, these combination tones become ever more dominant (Fig. \ref{fig:aplnova_fig6}, and Ref. \cite{MartignoliStoop2010}) and generate very complex responses that provide the substrate for the human-perceived pitch sensation \cite{MartignoliGomezStoop2013,Gomez_Stoop2014}.

\begin{figure}
\centering
\includegraphics[width=1\linewidth]{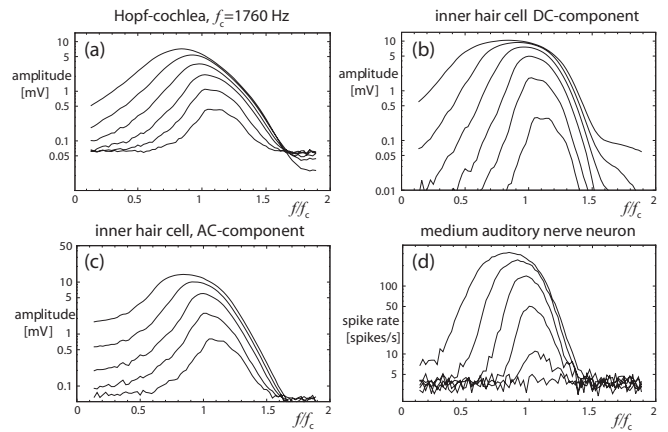}
\caption{Mammalian hearing along the auditory pathway a)-d), at a chosen frequency `channel'. Vertical direction describes amplification characteristics, horizontal direction expresses frequency-tuning; lines refer to equal input levels. At the end of the pathway (d), the cochlear sound information is practically unchanged \cite{MartignoliGomezStoop2013} (analog cochlea implementation \cite{stoopcochlea1,stoopcochlea2,vyver}).}
\label{fig:aplnova_fig3}
\end{figure}

Placing the frequency-specific sensors onto one area and arranging it in a tonotopic fashion, requires the animal to cope with the emergent interaction complexity. Astonishingly, the mammalian auditory system does not make any noticeable effort to correct for the (seen from the the classical signal processing dogma) undesired information gathered at the level of the cochlea.
Whereas filtering out at least some of the `artificial' components would certainly have been possible, this is just not how the mammalian hearing system works \cite{Abel}. In fact, biologically detailed simulations of the auditory pathway demonstrate that all the data collected at the cochlear level (including all interactions products) are as faithfully as possible transported along the pathway (c.f. Fig. \ref{fig:aplnova_fig3}), despite undergoing a whole astonishing variety of transformations and transductions on this way \cite{MartignoliGomezStoop2013}. This supports that pitch is already present at the cochlear level and is not primarily a cortical product \cite{MartignoliStoop2010,MartignoliGomezStoop2013,Gomez_Stoop2014}. In fact, we explicitly showed \cite{Gomez_Stoop2014} that the pitch extracted from the continuous physics at cochlear level fully coincides with the pitch extracted at the end of the auditory nerve from discrete spikes \cite{Rhode1995}.

\section{Coping with response complexity} 
The generated complex response along the cochlear duct even for simple stimulations  (Fig. \ref{fig:aplnova_fig6}), begs the question how the nervous system could cope with such a complexity. From other fields of physics we know that a common recipe to get a grip on emergent complexity is to attach an overall characterization to it (fractal dimensions or  Lyapunov exponents describe, e.g., the complexity generated by a chaotic process confined to a strange attractor). We put forward that a similar effect could be the deeper nature of pitch perception.

\begin{figure}[hhhht]
\centering
\includegraphics[width=1\linewidth]{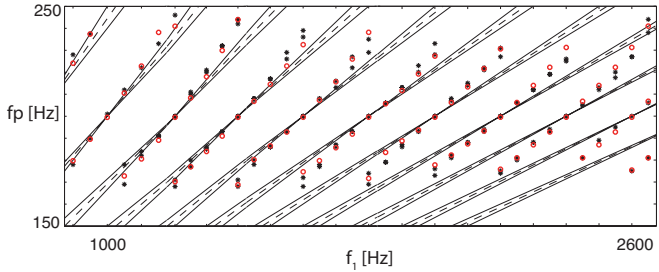}
\caption{Perceived pitch $fp$. For two-tone stimuli ($f_2=f_1+200$), Smoorenburg showed by his psychoacoustic experiments that the classically predicted pitch-shift of $\delta f/(k+1/2)$ does not emerge  (`second pitch shift': black dots (partial sound levels 40 dB sound pressure level, two subjects) vs. black lines \cite{Smoorenburg}). 
Red dots: Pitch extracted from a Hopf cochlea 
\cite{Gomez_Stoop2014}.
}
\label{fig:aplnova_fig7}
\end{figure}

In the simple case of pure tone stimulations, pitch sensation coincides with the obvious physical properties of the stimulator. For slightly more complicated stimulations, the generated response develops, however, a profile of its own that departs substantially from the physical properties of the stimulating signal, due to characteristics that are rooted in the interaction among the nonlinear sensors. Such is the origin of the celebrated second pitch shift (Fig. \ref{fig:aplnova_fig7}) investigated by Smoorenburg. 
Motivated by the {\it missing fundamental} paradigm, Smoorenburg performed psychoacoustical two-tone pitch-shift experiments. In these experiments, the perceived pitch regarding  an input of the form  $F_1 e^{2\pi i f_1 t} +F_2 e^{2 \pi i (f_1+200)t}$ was evaluated by well-trained subjects, and compared to what the then known physical theories would predict \cite{Smoorenburg}. The human result was found to, first, depart from what would have been expected from a `lowest order' stipulated `fundamental frequency' (first pitch shift), and, second (the second pitch shift), also differed when the emergence of combination tones was taken in a somewhat hand-waving way into account (de Boers's formula \cite{deBoer1956}). The psychoacoustic experiments manifest by up to three different perceived pitches for the same experiment. For what is actually heard as pitch, also the efferent connections to the cochlea seem to play a significant role and, along with this, earlier perceived sound \cite{Stoop_GomezREAPPL}.  
When the pitch was read out from our detailed model of the mammalian cochlea taking into account Smoorenburg's psychoacoustic and biophysical observations, the perceived pitch $f_p$ could be computed from the dominant peaks of the signal's
autocorrelation function and is found to fully agree with the psychophysical evaluations (the two- or even threefold ambiguity of pitch is a coherent observation in all these experiments  \cite{Gomez_Stoop2014}). The second pitch shift could then be attributed to the fact that the sound waves are transmitted by the cochlear fluid, an influence that previous theories of the perceived pitch entirely disregarded. Note that the wave-form of this signal is for complex sounds very different from the wave-form of the stimulating signal before entering the cochlea.

A `pitch sensation' tool seems thus to have been created very early in evolution, to cope with the generated signal complexity in the ear and to render a `purification' of the signal unnecessary. This might have been found to work much better than what more classical signal processing methods would ever be able to offer. The pitch sensation as defined jointly in terms of physics and physiology \cite{Gomez_Stoop2014} permits the auditory system to identify or tag even an inharmonic sound by condensed information as a``fundamental frequency'', even though the latter may be absent in the physical stimulus. This embracing property of pitch has recently been used as the main guiding principle in an approach \cite{Stoop_GomezREAPPL} for extracting elements of the auditory scene, which is at the heart of the cocktail party problem. Our physical understanding of the compound hearing sensor, the cochlea, also put forward how such a tuning in to a desired sound could be achieved by means of the efferent connections that exist from diverse levels of the auditory pathway to the cochlea. By assigning to these connections the task of tuning the amplifiers away from being effective (i.e., by moving the Hopf parameter further away from bifurcation) for undesired signals, using the pitch of the desired signal as the guiding feature, we can show how it is possible in a computationally cheap manner to extract desired and suppress undesired sounds, even if they partially overlap \cite{Stoop_GomezREAPPL}. In biological hearing, such a guidance may exploit past experience (by memory, we know what an instrument/speaker will sound like), or may exploit particularities of the signal extracted during the listening process. It is every-day experience that we invest considerable efforts for ``tuning in'' to a target sound, before we are able to follow it. 

\section{Conclusions} 
Our evidence from fundamental nonlinear physics, supports and explains the earlier observed convergence in hearing sensor construction. It complements the physiological and genetical findings
in a true sense, by reaching out towards the question why (instead of how) this convergence happened.
We expect that future studies from multiple approaches, of the hearing system as an evolutionary prototype will shed light on one fundamentally important question: What is the best computational framework for processing complex neural information? In light of the understanding that we have achieved regarding the first steps of the hearing pathway, this expectation does not appear to be overly optimistic. If so, then the physical principles underlying the hearing sensor evolution, would not only have provided us with the blueprints for artificial hearing sensors of power and abilities that match or even top the biological example, but would moreover reveal the principles followed by optimal signal processing structures.

\end{document}